# Current-induced nanogap formation and graphitization in boron-doped diamond films

V. Seshan,[1,2,+] C. R. Arroyo,[1,+,*] A. Castellanos-Gomez,[1,*] F. Prins,[1] M. L. Perrin,[1] S.D. Janssens,[3,4] K. Haenen,[3,4] M. Nesládek,[3,4] E.J.R. Sudhölter,[2] L.C.P.M. de Smet,[2] H.S.J. van der Zant,[1] and D. Dulic[1]

A high-current annealing technique is used to fabricate nanogaps and hybrid diamond/graphite structures in boron-doped nanocrystalline diamond films. Nanometer-sized gaps down to ~1 nm are produced using a feedback-controlled current annealing procedure. The nanogaps are characterized using scanning electron microscopy and electronic transport measurements. The structural changes produced by the elevated temperature, achieved by Joule heating during current annealing, are characterized using Raman spectroscopy. The formation of hybridized diamond/graphite structure is observed at the point of maximum heat accumulation.

During the last decade, carbon-based materials like diamond,[1] graphene[2] and carbon nanotubes (CNTs)[3] have been extensively studied due to their outstanding physical properties and potential applications. Consequently, hybrid structures combining the advantages of different allotropes of carbon into one structure have slowly gained interest. For instance, hybrid diamond/graphite structures can be relevant in molecular electronic applications because they benefit from both the robustness of

[1] Kavli Institute of Nanoscience, Delft University of Technology, Lorentzweg 1, 2628 CJ Delft, The Netherlands.
[2] Department of Chemical Engineering, Delft University of Technology, Julianalaan 136, 2628 BL Delft, The Netherlands
[3] Institute for Materials Research, Hasselt University, Wetenschapspark 1, BE-3590 Diepenbeek, Belgium
[4] IMOMEC, IMEC vzw, Wetenschapspark 1, BE-3590 Diepenbeek, Belgium
[+] Both authors have contributed equally.
[*] Corresponding authors: c.arroyorodriguez@tudelft.nl and a.castellanosgomez@tudelft.nl

$sp^3$ bonds and the flexibility of $sp^2$ hybridization for functionalizing with a large variety of molecules. Hybrid structures such as diamond/graphite nanowires,[4] diamond/graphite nanoflakes[5] and diamond/CNTs[6] composites have been reported in recent years.

So far these hybrid structures have been synthesized using the conventional plasma-based chemical vapor deposition (CVD) technique. This technique, however, does not allow *in-situ* fabrication of hybrid structures in specific domains. The current annealing technique,[7] on the other hand, has proven to be effective to induce structural transformations and even to fabricate nanometer-sized gaps. For example, the transformation of amorphous carbon layers into graphene[8] and restructuring of CNTs[9] have been reported using this technique. Furthermore, high-current annealing has also been used to fabricate nanogaps in few-layer graphene[10] and CNTs.[11] High-current annealing can therefore be a promising technique to fabricate nanogaps and hybrid structures also in diamond-based devices, which has not been explored yet. Nanocrystalline diamond[12] films are unique due to their close resemblance to single crystal diamond for many properties, the flexibility to grow on different substrates and the control over their electrical properties *via* boron doping (ranging from wide bandgap insulator to semiconductor to superconductor behavior). Boron-doped nanocrystalline diamond (B:NCD) films have been implemented as an electrochemical electrode,[13] sensor[14] and even superconducting quantum interference device (SQUID).[15] However, engineering B:NCD for using it as an electrode material for molecular electronics applications has been a big challenge because the strong covalent carbon network of diamond requires an unconventional approach to structure it.

In this letter, we introduce a technique involving current-induced annealing to create nanogaps as well as hybrid B:NCD/graphite structures. Using a feedback control loop, current annealing can be used to create gaps down to ~1 nm. The structural change from diamond to a graphitic phase in B:NCD produced by the Joule heating during the current annealing is characterized by Raman spectroscopy. The results presented here offer a method to engineer diamond devices to incorporate hybrid diamond/graphitic structures and nanogaps which cannot be performed using alternative techniques.[4-6]

B:NCD films were grown on a silicon oxide substrate in an *ASTeX* 6500 series using microwave plasma-enhanced CVD technique. Prior to the CVD, a silicon substrate was cleaned with nitric acid in an ultrasonic bath for 2-3 min and then rinsed with deionized water. The substrate was then dip-coated in a colloidal solution containing detonation diamond with a particle size of ~5-10 nm.[16] The resulting layer of diamond particles on the substrate acts as a nucleation center. A thin film was grown from the seeded substrate inside a CVD reactor using $H_2/CH_4$ plasma with a methane concentration around 3 vol%. The substrate temperature was maintained at 700 °C with process pressure of ~40 mbar and microwave power of 3500 W. Boron doping in the film was achieved by introducing trimethylboron (concentration ~3000 ppm) gas during the CVD process.[17]

B:NCD-based devices were fabricated from the film by using standard electron beam lithography. First, the B:NCD film was spin-coated with a double layer of resist, *i.e.* 330 nm of methyl-methacrylate (MMA, 17.5 %) and methacrylic acid (MAA, 8 %) copolymer followed by 250 nm of poly-MMA with a molecular weight of 950 kD and then prebaked at 175 °C for ~12 min. The resists were exposed to electron beam with a dose of 900 $\mu C/cm^2$ and an acceleration voltage of 100 kV. The development of the exposed areas was done by dipping the substrate in a 1:3 volume of methyl isobutyl ketone (MIBK) and isopropyl alcohol (IPA) for 2 min followed by IPA for 1 min. A thin layer of aluminium (~60 nm thick) was deposited on the developed part of resist by electron beam evaporation followed by lift-off in hot acetone. The thin film was then etched using oxygen ($O_2$) reactive ion etching (RIE) with a DC bias voltage of –413 V, $O_2$ gas flow of 30 ml/min, pressure of 20.7 μbar and power of 30 W for ~22 min. The aluminium film acts as a mask during the etching process thereby protecting the diamond film/structures underneath. After the RIE process, the protective aluminium layer was removed using 0.8 wt% KOH solution. In a second lithographic step, contact pads of 20 nm titanium/200 nm gold were fabricated. In order to fully characterize the electron transport properties and nanogaps in B:NCD film, two batches of devices were fabricated. One batch consisted of Hall bar geometry while the second batch consisted of 2-terminal geometry. After fabrication, the devices were characterized employing optical microscopy, atomic force microscopy (AFM), electrical transport measurements at low temperature, scanning electron microscopy (SEM) and Raman spectroscopy.

Figure 1(a) shows an optical micrograph using a 100× magnifying objective and (b) shows the corresponding SEM image of a device fabricated following the steps described above. The thickness of this B:NCD film/device has been measured by means of AFM (scanned area marked with the dashed square in Fig. 1(a)) and was found to be ~180 nm (Fig. 1(c)). The Hall bar devices were electrically characterized using a 4-terminal geometry in a $^3$He/$^4$He dilution refrigerator with a base temperature of 20 mK by injecting a DC current and measuring the voltage drop across two electrodes. Figure 1(d) presents the current *vs.* voltage characteristics for a device at 5.4 K (red curve) and 54 mK (blue curve). While at 5.4 K the device behaves as a normal ohmic conductor, at 54 mK the transport characteristics show signs of superconducting behavior as reported earlier for this material.[18] For this device, the critical current was found to be ~100 nA. The onset of superconductivity for our B:NCD devices was found to be between 1 and 2 K.

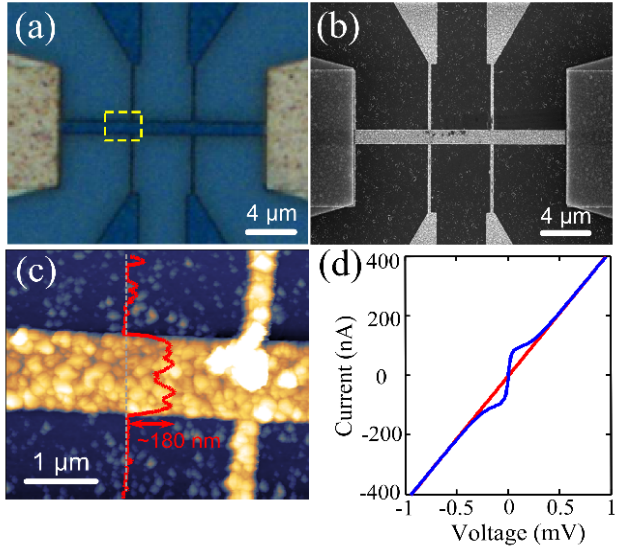

FIG. 1. (a) Optical image using a 100× magnifying objective marked with AFM scan area (dashed square) and (b) corresponding SEM image of the device fabricated on the B:NCD film. (c) Topography AFM image of the device with height profile (red curve). (d) Current *vs.* voltage characteristics of the device measured at 5.4 K (red curve) and 54 mK (blue curve).

B:NCD devices were then subjected to high-current annealing in air at room temperature using a feedback-controlled scheme, as shown in Fig. 2(a). Although the devices were fabricated in Hall bar geometry, for current annealing they were used in 2-terminal geometry. A voltage (*V*) ramp with a rate of 1 V/s was applied to the device while the current (*I*) was measured every 100 μs. If the conductance (*I*/*V*) drops >10 % within a 200 mV range, the voltage was swept back to zero after which a new ramp was started. This process was repeated until a gap was formed (Fig. 2(b)). Figure 2(c) shows a typical *I-V* graph of the feedback-controlled high-current annealing process. The current increases with the voltage until a critical point where the conductance starts to decrease (indicated by the red circles in Fig. 2(c) for easy detection). The arrow in Fig. 2(c) indicates the evolution of the *I-V* traces after sequential high-current annealing steps, eventually leading to a nanogap. The nanogaps in the first batch of B:NCD devices were found to be around ~100 nm and approximately in the middle part between the two leads.

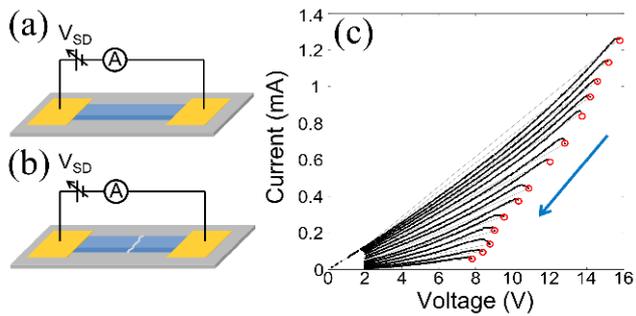

FIG. 2. Schematic diagram of the feedback-controlled, high-current annealing process, before (a) and after (b) the formation of a nanometer-sized gap in the B:NCD device. (c) *I-V* graph of feedback-controlled, high-current annealing process. The red circles (for easy detection) indicate decrease in conductance while the arrow indicates the evolution of the *I-V* traces after sequential high-current annealing steps leading to a nanogap eventually.

During the high-current annealing, the film can experience heating which may lead to structural changes. In order to gain a deeper insight into the structural changes induced by the high-current annealing, Raman spectroscopy has been employed. This technique has been extensively used to study different allotropic forms of carbon-based materials.[19,20] A micro-Raman spectrometer (*Renishaw in via RM 2000*) was used in a backscattering configuration excited with visible laser light ($\lambda$ = 514 nm). The spectra were collected through a 100× magnifying objective and recorded with 1800 lines/mm grating providing the spectral resolution of ~1 cm$^{-1}$. Before the high-current annealing, the B:NCD devices show all the typical features of boron-doped nanocrystalline diamond, *i.e.* peaks around 1300 and 1590 cm$^{-1}$, ascribed to the diamond and *G* band, in addition to another peak around 1210 cm$^{-1}$ attributed to boron doping in diamond (Fig. 3(a)).[21,22] The presence of a marginal *G* peak in the Raman spectrum is due to the presence of graphitic phases at the grain boundaries. The peak below 1000 cm$^{-1}$ corresponds to the excitation of a second order phonon in the silicon which was used as the substrate to grow B:NCD film.[21]

After the high-current annealing, on the other hand, the Raman spectrum measured close to the nanogap changes drastically (Fig. 3(a)). The peak at 1590 cm$^{-1}$ is more pronounced after current annealing and a new peak at 1347 cm$^{-1}$ (*D* peak) is present, typically associated with the presence of sp$^2$ hybridized carbon.[19] Therefore, the Raman spectra after the current annealing and near the nanogap resemble that of a graphitic phase, dominated by sp$^2$ instead of sp$^3$ hybridization. Such features have previously been observed for graphitized amorphous carbon samples subjected to high temperature annealing, where the appearance of these features was attributed to a conversion from sp$^3$ to sp$^2$ hybridization.[19] In the case of current-induced annealing, the B:NCD samples experience Joule heating due to large current flow through the device (~$1.2 \times 10^7$ A/cm$^2$). The highest temperature is reached in the central part between the drain and source leads since the (gold) electrodes/contacts act as a heat-sink. Indeed, the graphitic-like Raman spectrum is not observed in the close vicinity of the electrodes where the increase of temperature is lower due to efficient heat dissipation through the electrodes.

The spatial distribution of the structural changes induced by high-current annealing has been studied by performing scanning Raman spectroscopy. The inset in Fig. 3(b) shows an optical image of a B:NCD device which has been subjected to the current annealing procedure resulting in a wide gap (~100 nm). Scanning Raman spectroscopy (laser spot size ~400 nm) has been carried out in the region highlighted with the dashed square. Figures 3(b)-(d) shows the intensity profile of the Raman peak at 1211 cm$^{-1}$ (boron peak), 1347 cm$^{-1}$ (*D* peak) and 1590 cm$^{-1}$ (*G* peak) respectively. From these figures, one can conclude that the graphitic phases are located around the gap, indicating that the temperature at that spot reached an elevated temperature during the current annealing. Notice that similar graphitization processes have been achieved by annealing amorphous carbon[19] and nanodiamonds[23] around 1000 °C. This gives an estimation of the temperature reached in our device during our current annealing. The boron peak, on the other hand, shows a homogeneous intensity all along the device except for the gap region where the material has been removed by the current annealing process. However, due to the laser spot size (~400 nm) and the possible light scattering close to the gap edges, one cannot characterize the structures below 200 nm and thus the gap cannot be resolved by scanning Raman spectroscopy. Interestingly, we have observed that, due to the device geometry, some other parts can experience high temperature during the current annealing but without reaching the break point, which shows up as graphitic-like Raman spectrum in some regions far from the gap (see the right side of the central bar in Fig. 3(c) and (d)).

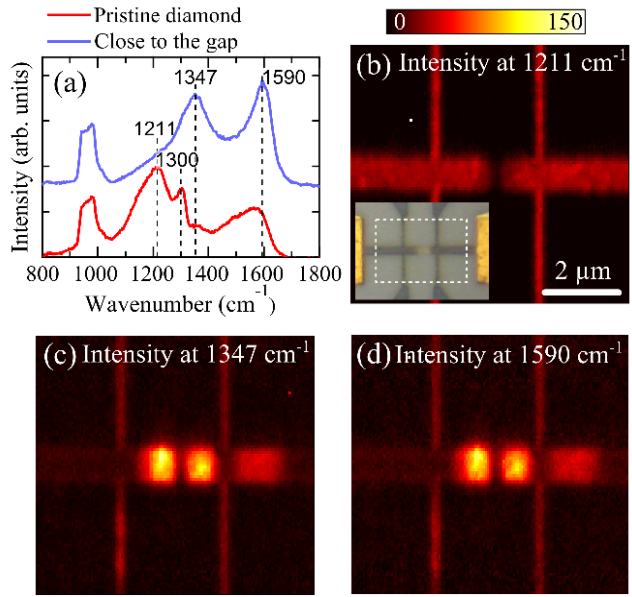

FIG. 3. (a) Raman spectra of the pristine B:NCD film (red curve) and close to the gap (blue curve) after high-current annealing process. Spatial map of the intensity of the Raman peak at 1211 cm$^{-1}$ (b) with inset showing the laser scanning area (white dashed square) on the device, 1347 cm$^{-1}$ (c) and 1590 cm$^{-1}$ (d).

While for the first batch of B:NCD devices the gap size was on the ~100 nm range, a second batch of devices, with 2-terminal geometry and smaller cross section area ($100 \times 100$ nm$^2$), yielded more controlled, high-current annealing process and smaller gaps in the ~1 − 10 nm range. Overall, we have performed current annealing on 19 devices of which 15 devices showed low-bias resistance in the range of 30 MΩ - 30 GΩ while rest had infinite resistance (>100 GΩ). After the current-induced annealing, the samples have been inspected using SEM (*Hitachi high resolution FE-SEM (S-4800)*). Figure 4(a) shows a close up SEM image of a nanogap. The SEM image illustrates how the nanogap is "wedge" shaped, being narrower in the bottom part where electrodes are very close (~ nm range) and the tunneling process takes place. Quantitative analysis of top-view and high-angle SEM images gives an estimate of the tunneling area ~$30 \times 30$ nm$^2$. Figure 4(b) shows a typical tunneling *I-V* trace (red curve) measured at room temperature across one of the 15 devices annealed until opening a narrow nanogap/junction (see the supporting information for temperature dependence). The tunneling phenomenon is an indication that our gap size is in the order of ~1-2 nm as estimated using the Simmons model[24] (fit shown as black curve in Fig. 4(b)). The parameters used in the fit are for the area (*A*): 900 nm$^2$ (~$30 \times 30$ nm$^2$), barrier height ($\phi$): 1.15 eV and gap size (*z*): 1.42 nm. Although the barrier height is expected to be similar to the work function (~4 eV), for electron tunneling under ambient conditions, lower values (around 1 eV) have been observed for other carbon-based materials such as graphene nanogaps[10] and carbon fiber-based scanning tunneling tips.[25] This procedure, therefore, can be used to fabricate nanometer-spaced electrodes and to engineer hybrid diamond/graphitic structures at specific places in the device.

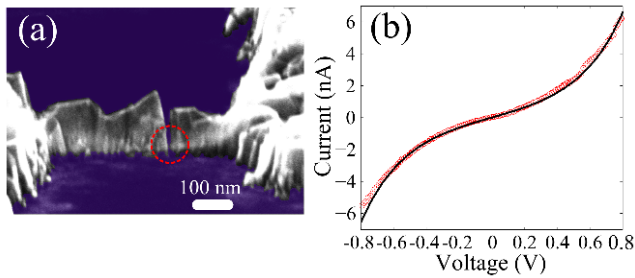

FIG. 4. (a) Higher magnification SEM image (with false colored silicon oxide substrate) of one of the fabricated gaps in the B:NCD device. The red dotted circle indicates possible tunneling region. (b) *I-V* characteristics across the nanogap (red curve) and corresponding fit using the Simmons model for tunneling (black curve). Fit parameters used in Simmons model are area ($A$): 900 nm$^2$, barrier height ($\phi$): 1.15 eV and gap size ($z$): 1.42 nm.

We present a study on the current-induced annealing in boron-doped nanocrystalline diamond devices. Using a feedback control loop, the current annealing technique can be used to create nanosized gaps down to ~1 nm. The structural changes produced by the current-induced heating have been studied by Raman spectroscopy. We found that Joule heating increases the temperature enough to change the hybridization of diamond, dominated by sp$^3$, to a graphitic sp$^2$ hybridized. The study reported here presents a way to engineer diamond-based devices into hybrid diamond/graphitic structures, with prospective use as nanoelectrodes, for instance in the field of molecular electronics.

## ACKNOWLEDGMENTS

We thank B. Schneider for SEM imaging. This work was supported by the Delft University of Technology, Hasselt University, the Dutch organization for Fundamental Research on Matter (FOM), the Research Programs G.0555.10N of the Research Foundation-Flanders (FWO), the European Union (FP7) through the programs RODIN, MOLESOL, DIAMANT (Co. N 270197) and the Marie Curie ITN "MATCON" (PITN-GA-2009-238201).

# Supporting information for
# Current-induced nanogap formation and graphitization in boron-doped diamond films

V. Seshan,[1,2,a)] C. R. Arroyo,[1,a),b)] A. Castellanos-Gomez,[1,c)] F. Prins,[1] M. L. Perrin,[1] S.D. Janssens,[3,4] K. Haenen,[3,4] M. Nesládek,[3,4] E.J.R. Sudhölter,[2] L.C.P.M. de Smet,[2] H.S.J. van der Zant,[1] and D. Dulic[1]

[1]Kavli Institute of Nanoscience, Delft University of Technology, Lorentzweg 1, 2628 CJ Delft, The Netherlands

[2]Department of Chemical Engineering, Delft University of Technology, Julianalaan 136, 2628 BL Delft, The Netherlands

[3]Institute for Materials Research, Hasselt University, Wetenschapspark 1, BE-3590 Diepenbeek, Belgium

[4]IMOMEC, IMEC vzw, Wetenschapspark 1, BE-3590 Diepenbeek, Belgium

[a)]V. Seshan and C. R. Arroyo contributed equally to this work.

[b)]Electronic mail: c.arroyorodriguez@tudelft.nl.

[c)]Electronic mail: a.castellanosgomez@tudelft.nl.

**Temperature dependence of the tunneling characteristics of the diamond gaps**

We have investigated the nature of the electron transport in the diamond devices with nanogaps by recording current versus voltage (IV) characteristics at different temperatures. Figure S1 shows the temperature dependence IV plot for one of the devices with a nanogap. The IV characteristics look similar at all the temperatures except for slight decrease in current at lower temperatures. The observed differences in the IV characteristics are consistent with tunneling behavior as the main transport mechanism through the gap.[1]

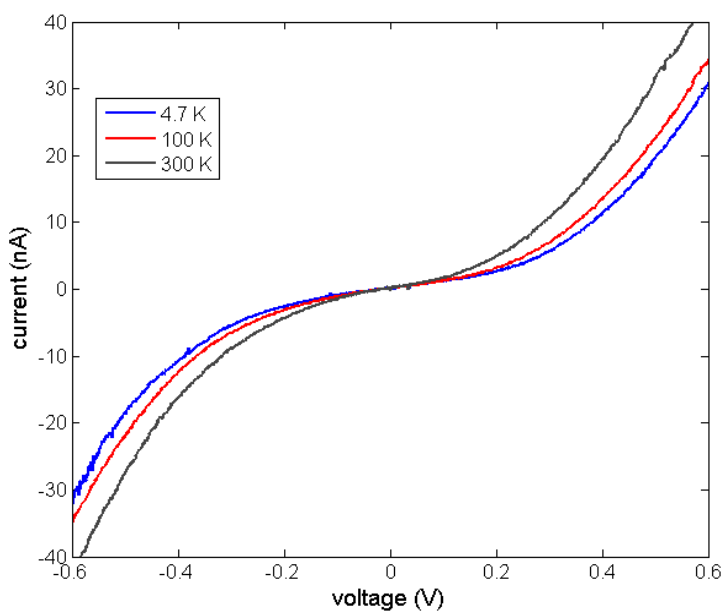

Figure S1: Current versus voltage characteristics of a diamond device with a nanogap at different temperatures.

**Reference**

[1] J. G. Simmons, J. Appl. Phys. **34,** 6 (1963).